\documentclass[12pt,preprint]{aastex}

\begin{document}

\title{A new method to estimate local pitch angles in spiral galaxies:
Application to spiral arms and feathers in M81 and M51}

\author{Iv\^anio Puerari\altaffilmark{1}, Bruce G. Elmegreen\altaffilmark{2}, David L. Block\altaffilmark{3}}

\altaffiltext{1}{Instituto Nacional de Astrof\'\i sica, Optica y Electr\'onica, Calle Luis Enrique Erro 1, 72840 Santa Mar\'\i a Tonantzintla, Puebla, Mexico (puerari@inaoep.mx)}
\altaffiltext{2}{IBM T. J. Watson Research Center, 1101 Kitchawan Road, Yorktown Heights, New York 10598 USA}
\altaffiltext{3}{School of Computational and Applied Mathematics, University of Witwatersrand, Private Bag 3, WITS 2050, South Africa}

\begin{abstract}
We examine $8\mu$m IRAC images of the grand design two-arm spiral galaxies
M81 and M51 using a new method whereby pitch angles are locally determined
as a function of scale and position, in contrast to traditional Fourier
transform spectral analyses which fit to average pitch angles for whole
galaxies.  The new analysis is based on a correlation between pieces of a
galaxy in circular windows of $(\ln R , \theta)$ space and logarithmic
spirals with various pitch angles. The diameter of the windows is varied to
study different scales. The result is a best-fit pitch angle to the spiral
structure as a function of position and scale, or a distribution function
of pitch angles as a function of scale for a given galactic region or area.
We apply the method to determine the distribution of pitch angles in the
arm and interarm regions of these two galaxies.  In the arms, the method
reproduces the known pitch angles for the main spirals on a large scale,
but also shows higher pitch angles on smaller scales resulting from dust
feathers. For the interarms, there is a broad distribution of pitch angles
representing the continuation and evolution of the spiral arm feathers as
the flow moves into the interarm regions.  Our method shows a multiplicity
of spiral structures on different scales, as expected from gas flow
processes in a gravitating, turbulent and shearing interstellar medium. We
also present results for M81 using classical 1D and 2D Fourier transforms,
together with a new correlation method, which shows good agreement with
conventional 2D Fourier transforms.
\end{abstract}

\section{Introduction}
\label{introduction}

Disk galaxies support stellar spiral density waves \citep{bertinetal89a}
that interact with each other at resonances \citep{tagger87} to produce
complex, regenerating structures \citep[see review in][]{sellwood2014}.
These waves also interact with the gas to produce even more complex
structures, including gas clumps in the arms \citep{kim01}, gas feathers
downstream from the main arms \citep{kimostriker02}, and multiple shocks 
\citep{lugovskii14}. These and other gas structures have further
connections with star formation, which can produce its own
structure, such as shells, pillars, and local blowout cavities
\citep[see review in][]{elmegreen12}.  The quantification of all this
structure for the purpose of comparing different regions or different
galaxies, or comparing theory with observations, is difficult because the
structure is multi-scale, multi-component (gas, old stars, young stars,
etc.), and spatially varying.

Here we describe a new method for quantifying spiral structure and its
spatial variations covering a wide range of scales.  We apply this method
to the characterization of spiral arms and feathers for two grand design
galaxies, M81 and M51.  The $8\mu$m images from Spitzer are used because
they show essentially all of the diffuse gas structures in PAH emission
with high angular resolution; $0.748^{\prime\prime}$ for M81, which is 13
pc for a distance of 3.6 Mpc, and $1.222^{\prime\prime}$ for M51, which is
63 pc for a distance of 10.6 Mpc (NASA/IPAC Extragalactic Database,
http://ned.ipac.caltech.edu).


\cite{sandage1961} noted the prominence of dust lanes in late-type spiral
galaxies, describing stellar ``branches'' in the main arms and dust
``filaments'' that cut across the arms, especially in the case of M51.
\cite{lynds1970} studied 17 late-type spirals using archival photographic
plates from the Mount Wilson and Palomar observatories. She also noted the
presence of main dust lanes in the arms and thin ``feathers'' of dust with
large pitch angles cutting across the arms. \cite{weaver70} placed the sun
in a stellar ``spur'' or ``offshoot" of the Sagittarius arm in the Milky
Way. \cite{elmegreen80} studied seven spirals to investigate in more detail
the properties of {\it stellar} spurs and determined that they have pitch
angles $\sim50^\circ$ larger than the main arms. She noted that gaseous
feathers and stellar spurs have similar pitch angles and suggested they may
have a common origin \citep[see also][]{piddington1973}. A large survey of
dust feathers in spiral galaxies was conducted by \cite{lavigneetal2006}
using archival data from the Hubble Space Telescope. They found a decrease
in spur separation with increasing gas density, suggesting that
gravitational instabilities are involved.  Periodic dust feathers occurred
in 20\% of their Sb and Sc types.

In what follows, we define ``spurs'' as interarm stellar features that jut
out from otherwise continuous stellar arms, as distinct from branches,
where one arm ends and turns into two arms further out.  Similarly, we
define ``feathers'' as gas or dust features that jut out from the main
stellar arms.  These terms are consistent with the usages referenced above.
The present study is about feathers because we use dust emission to
delineate the structures.

Interarm dust feathers were first seen in emission at $15\mu$m for M51
\citep{block97} with ISOCAM \citep{cesarsky96}. They were first seen at
arcsec resolution in M81 with Spitzer $8\mu$m images
\citep{willneretal2004}. Feathers are bright in molecular line emission as
well, and they often have associated H$\alpha$ \citep[for M51,
see][]{scoville01,corderetal2008,koda09,schinnereretal2013}.
\cite{chandar11} note that two prominent feathers in M51 contain clusters
$\sim10^8$ yrs old and suggest this is consistent with simulations of
global spiral density wave arms in \cite{dobbs10}.

Efforts to explain spurs and feathers began with \cite{balbus88}, who
performed a local gas dynamical stability analysis of a single-fluid
polytropic flow through a spiral arm. He found that gravitational
instabilities in the presence of reverse shear and expansion downstream
from the arm can be reinforced by epicyclic motions and lead to the growth
of features with large pitch angles. Two-dimensional, time-dependant,
magnetohydrodynamic simulations of a self-gravitating, differentially
rotating piece of a gas disk with a steady spiral potential were studied by
\cite{kimostriker02}. They also found that gravitational instabilities in
the compressed arm gas produced feathers downstream as a result of
expansion and reverse shear. Extension to three-dimensions confirmed these
results, although the separation between feathers increased because of the
dilution of the in-plane component of gravity \citep{kim06}.
\cite{chakrabartietal2003} studied the growth of small-scale gas features
in a steadily imposed stellar spiral wave, finding feathers, branches, and
other chaotic features with a link to ultraharmonic resonances and
gravitational instabilities in the arms.

\cite{wadakoda2004} discovered an additional ``wiggle'' instability that
operates even without self-gravity when the reverse shear downstream from
an arm triggers something like a Kelvin-Helmholtz instability. Magnetic
fields \citep{shetty06,dobbsprice08} and three-dimensional effects such as
vertical shear \citep{kim06} may stabilize this, however.
\cite{dwarkadas96} suggested that the radial component of the post-shock
flow stabilizes the Kelvin-Helmholtz instability. \cite{kim14} looked at
the wiggle instability again and showed analytically and with hydrodynamic
non-gravitating simulations that it can arise from accumulated potential
vorticity in gas that flows successively through many irregular shock
fronts. They also explained why \cite{dwarkadas96} did not see the
instability; i.e., the growth rate for their long wavelength perturbations
was too slow. \cite{dobbs06} also got interarm feathers without
self-gravity, from sheared and expanded gas structures that form randomly
in the cold gas of spiral arms and then flow into the interarm region.

\cite{shetty06} extended the \cite{kimostriker02} two-dimensional
self-gravitating simulations to global scales and confirmed that feathers
grow from perturbations in spiral shocks for the inner regions. In the
outer regions, they found that interarm filaments can grow by local
self-gravity even without an imposed spiral. \cite{wada08} included
realistic heating and cooling in a three-dimensional self-gravitating
hydrodynamics simulation with an imposed spiral and found feathers
connected with the main shocks. In contrast, \cite{wada11} got no spurs or
feathers in N-body+SPH models with live stars and gas because the spirals
corotated with the disk while forming and re-forming, and the gas fell into
the spiral potential from both sides. In all of these models, gas
irregularities in the flow through a spiral wave of stars is required for
feather formation downstream.

Recently, \cite{leeshu2012} and \cite{lee2013} present an analytic
formulation of two-dimensional feathering in a magnetic, self-gravitating
gas that circulates in a galaxy with a steady stellar spiral.  This
formulation involves perturbing the shocked flow in spiral coordinates. A
similar procedure was followed by \cite{elmegreen91} to determine the
growth rate and effective Toomre $Q$ for shearing gravitational
instabilities in a regular spiral density wave flow. Both studies suggest
that these instabilities drive the formation of giant molecular clouds and
star formation in spiral galaxies \citep[see
also][]{khoperskov13,elmegreenetal2014}.

\cite{kimkim14} studied angular momentum transfer and radial gas drift in
two-dimensional hydrodynamical simulations of gas flow in an imposed spiral
arm potential. They show the density distribution in the ($\ln R,\theta$)
plane, which is similar to our display here. Interarm feathers with large
pitch angles are clearly present in their work, and they are connected with
density irregularities in the spiral arm shocks.

Renaud et al. (2013) performed an N-body/adaptive mesh simulation of a
Milky Way-size galaxy at extremely high resolution and got curled spiral
arm structures that resembled Kelvin-Helmholtz instabilities. They made a
distinction between small-scale structures, which host star formation at
their tips, and extended lower-density interarm structures, which have no
star formation.

To provide a quantitative basis for these studies, we introduce a method to
determine the distribution of pitch angles for spiral arms and their
smaller scale substructures. Using M81 and M51 as examples (Section
\ref{thedata}), we first review the standard Fourier transform techniques
and introduce an improved method using correlations in $2\pi$ azimuthal
windows, which give results in agreement with the Fourier techniques
(Section \ref{1D2DFourier_and_correlation}).  Section \ref{new_correlation}
then presents a better method that calculates correlations using small
circular windows in $(\ln R,\theta)$ space. This gives local pitch angles
at different scales, depending on the window size. Finally, in Section
\ref{results}, we present the results of this new method for M81 and M51.

\section{The data}
\label{thedata}

Incorporating large-format infrared detector arrays, with the intrinsic
sensitivity of a cryogenically-cooled mirror and the high observing
efficiency of a heliocentric orbit, the Spitzer Space Telescope gives
unparalleled opportunities to study spiral galaxy morphology and dust grain
emission. The IRAC instrument \citep{fazioetal04} comprises four detectors,
which after launch in August 2003, operated  for several years at four
wavelengths, these being 3.6 $\mu$m (channel 1), 4.5 $\mu$m (channel 2),
5.8 $\mu$m (channel 3) and 8.0 $\mu$m (channel 4). The IRAC filter band
center is 7.87 $\mu$m in channel 4 \citep[see][] {gehrzetal07}.

For both M81 and M51, we selected IRAC channel 4, wherein one observes
emission from ultra-small ($\sim 0.01 \mu$m) dust grains as well as from
macromolecules (the 8 micron band contains emission from polycyclic
aromatic hydrocarbons (PAHs) at $\lambda$ = 7.7 microns).  Small
carbonaceous grains and PAHs undergo temperature spiking
\citep{greenberg1968,sellgren84,greenberg1996,li2004} and can become hotter
than 1000K - 2000K as they transiently absorb a photon from the
interstellar radiation field. It is primarily the emission from such warm
($\sim$ 60 K) tiny dust grains and PAHs, subject to temperature spiking,
which is detected at 8 microns. As emphasized by \cite{bendoetal2008}, PAH
$8\mu$m emission seems to appear in shell-like features around star-forming
regions, and is thus an excellent tracer of these. In contrast, longer
wavelength $24 \mu$m emission is shown by \cite{bendoetal2008} to peak
within star-forming regions. Channel 4 is thus our preferred choice to
trace spiral arm structure, as well as star formation, in the disks of M81
and M51. Our re-binned channel 4 images of M81 and M51 have scales of
$0.748^{\prime\prime}$ pixel$^{-1}$ and $1.222^{\prime\prime}$
pixel$^{-1}$, respectively.

\section{1D and 2D Fourier techniques \& a correlation method}
\label{1D2DFourier_and_correlation}

The most common way to quantify $m$-armed spiral structure in disk galaxies
is with 1D Fourier transforms of the azimuthal profiles \citep[see
e.g.][]{grosboletal2004}. Essentially, for each azimuthal profile
$I_R(\theta)$, the order-$m$ Fourier coefficients $A_m$ are calculated as
functions of radius $R$ using the equation
\begin{equation}
A_m(R)=\int_{-\pi}^{\pi} I_R(\theta)e^{-im\theta}d\theta .
\label{equation_1dft}
\end{equation}
The amplitude of $A_m(R)$ determines the radial region where a given $m$
structure is important; the phase of $A_m(R)$ can be used to estimate the
pitch angle of the order-$m$ spiral arm.

Bidimensional Fourier techniques have been discussed in a number of papers
\citep[e.g.,][amongst
others]{kalnajs75,consathan82,iyeetal82,krakowetal1982,pueraridottori92,
puerari93,davidivanio99,puerarietal00}. They give global pitch angles,
averaged over a galaxy, or average pitch angles as a function of
galactocentric radius \citep{savchenko2012,davisetal2012}. The
bidimensional Fourier coefficients $A(m,p)$ in a basis of logarithmic
spirals can be calculated as
\begin{equation}
A(m,p)=\int_{u_{min}}^{u_{max}} \int_{-\pi}^{\pi} I(u,\theta)e^{-i(m\theta+pu)}d\theta du
\label{equation_2dft}
\end{equation}
where $u=\ln R$, $m$ is the azimuthal frequency (related to $\theta$), and
$p$ is the frequency related to $\ln R$. The pitch angle $P$ of the
order-$m$ spiral arm is given by $\tan P=-m/p$. $I(u,\theta)$ is the light
distribution of the galaxy in a $(u,\theta)$ plane. Hence, once the radial
annulus to be analyzed is chosen by fixing $u_{min}$ and $u_{max}$, the
amplitude of the complex matrix $A(m,p)$ will show the most probable pitch
angle $P$ of that $m$ structure in the annulus.

We discuss now a new method to extract the same information as the
bidimensional Fourier transform. We construct a family of synthetic
logarithmic spirals inside a $2\pi$ window in $(\ln R, \theta)$
coordinates, i.e., with a given $\Delta \ln R$ and $\Delta \theta=2\pi$ for
radius $R$ and azimuthal angle $\theta$. The synthetic spirals have a pitch
angle $P$ and a number of arms $m$, so their curves of constant phase are
given by $\ln R={m\over p}\theta+\Phi$ for some constant $\Phi$. The pitch
angle $P$ of the spiral in this formalism is given by $\tan P={{\Delta
R\over R\Delta \theta}} = {{\Delta \ln R\over \Delta \theta}} = {-{m\over p}}$.

The phase of the spiral is defined to be zero, which corresponds to the
crest of the synthetic arm, at the center of the window, so that $\Phi=\ln
R_{\rm min}+0.5\Delta \ln R$ for minimum window radius $R_{\rm min}$. Then
the synthetic spiral, $S$, is the cosine function of the phase,
\begin{equation}
S_{m,p}(\ln R,\theta)=\cos(m\theta+p\ln R+p\ln R_{\rm mid})
\end{equation}

With this definition of $\Phi$, the synthetic arm peak is in the middle of the
window for all assumed $m$ and $p$.  Note that for $p>0$, the azimuthal angle
increases in the usual sense for cylindrical coordinates, counter-clockwise, as the
radius increases, corresponding to a ``Z''-type morphology, while for $p<0$, the
spiral has an ``S''-type morphology.

The deprojected image $I$ of a galaxy - now sampled in a $(\ln R, \theta)$
plane - is then cross-correlated with the synthetic logarithmic spiral,
giving the correlation
\begin{equation}
C_m(\ln R,\theta,p)=\sum_x \sum_y S_{m,p}(\ln R+y,\theta+x)I(\ln R,\theta)
\label{corr_method_1}
\end{equation}
where the summation is over pixels in the range $x=-\pi$ to $\pi$ and
$y=\ln R-0.5\Delta \ln R$ to $\ln R+0.5\Delta \ln R$. This is the sum over
the local $(x,y)$ coordinates inside the window.  There is a different
correlation sum for each image coordinate $(\ln R,\theta)$, and for each
$p$ and $m$, i.e., $C$ is a 4 dimensional matrix. The pitch angle for an
$m$ structure at a given radius $\ln R_{\rm mid}$, is calculated using the
$p$ of the the maximum $C_m(\ln R_{mid},\theta, p)$, maximized in $\theta$.
This corresponds to the best-fit pitch angle for an $m-$arm spiral going
through the image coordinate $(\ln R_{mid},\theta)$.

\begin{figure}
\centering
\includegraphics[width=6.0in]{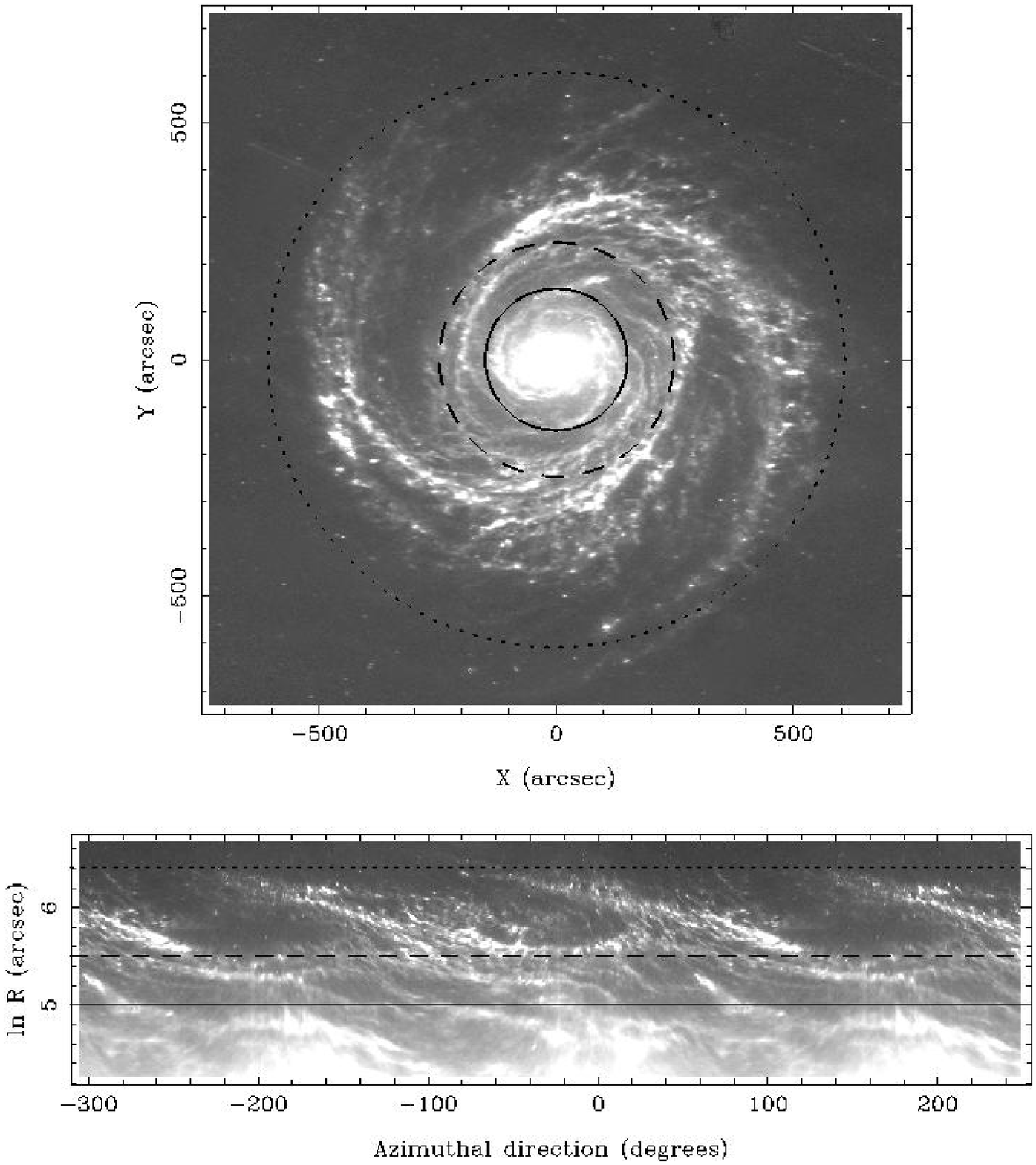}
\caption{M81 deprojected image in $(x,y)$ and in $(\ln R,
\theta)$. The circles in the upper panel and the horizontal lines in the
bottom panel correspond to $R=$ 150, 247 and 606 arcsec, or $\ln R=$ 5.0,
5.5, and 6.4. These radii are illustrative, and were taken from Figure
\ref{1dft}. The first value is at a minimum in the strength of the internal
$m=2$ structure, while the second is the radial position where
the main long external $m=2$ structure begins. The large radius is
where the amplitude of the $m=2$ structure decreases to 10\%
of its maximum value.} \label{galaxy_xy_lrt}
\end{figure}

\begin{figure}
\centering
\includegraphics[width=6.0in]{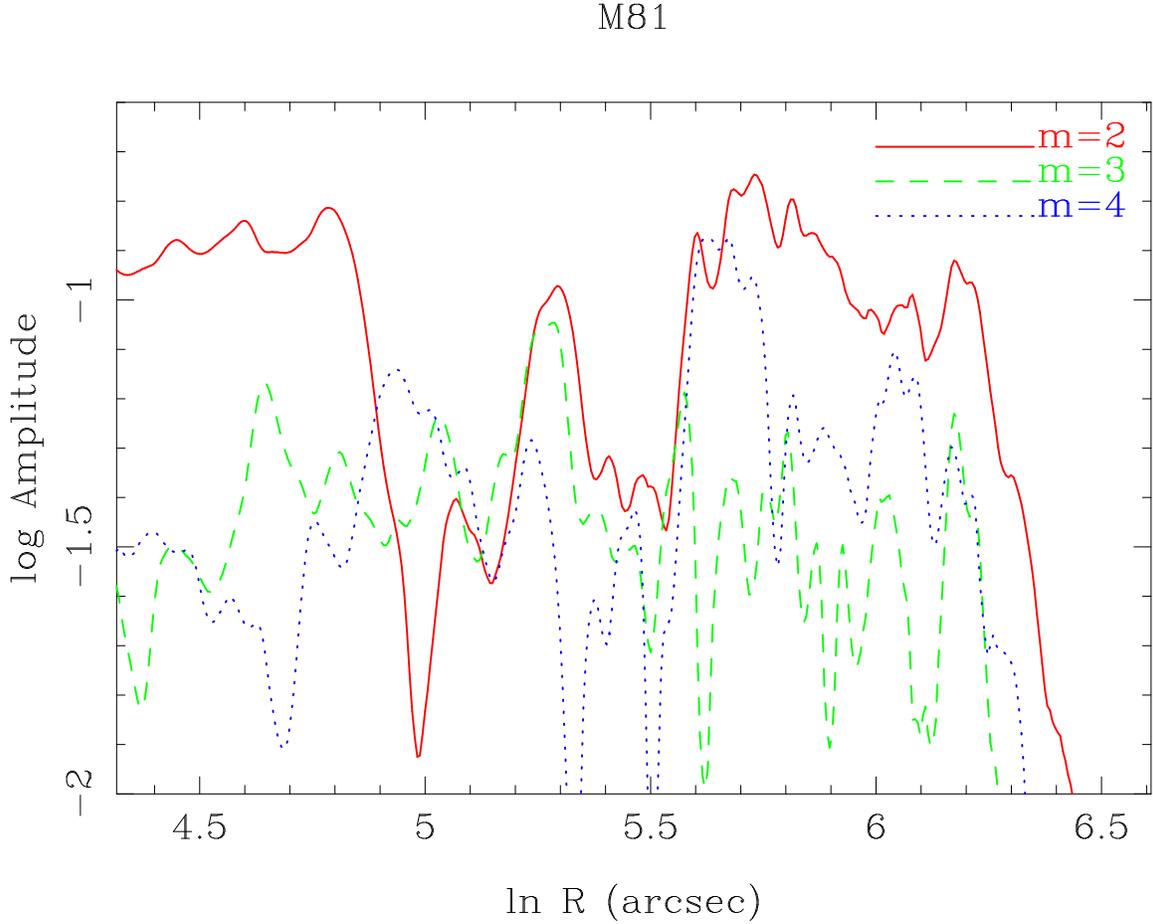}
\caption{Amplitudes of the 1D Fourier transform as a function of $\ln R$
for $m=$2, 3, and 4. Note the valleys in the $m=2$ curve at $\ln R=$ 5.0,
5.5, and 6.4. The corresponding radii (150, 247 and 606 arcsec) are
displayed in Figure \ref{galaxy_xy_lrt} (upper panel). Note that the amplitude for $m=2$
at $\ln R=5.0$ is only 1/3 of the maximum amplitude. For radii larger than
$\ln R=6.4$, the amplitude of the $m=2$ component is less than 10\% of its
maximum value.} \label{1dft}
\end{figure}

\begin{figure}
\centering
\includegraphics[width=6.0in]{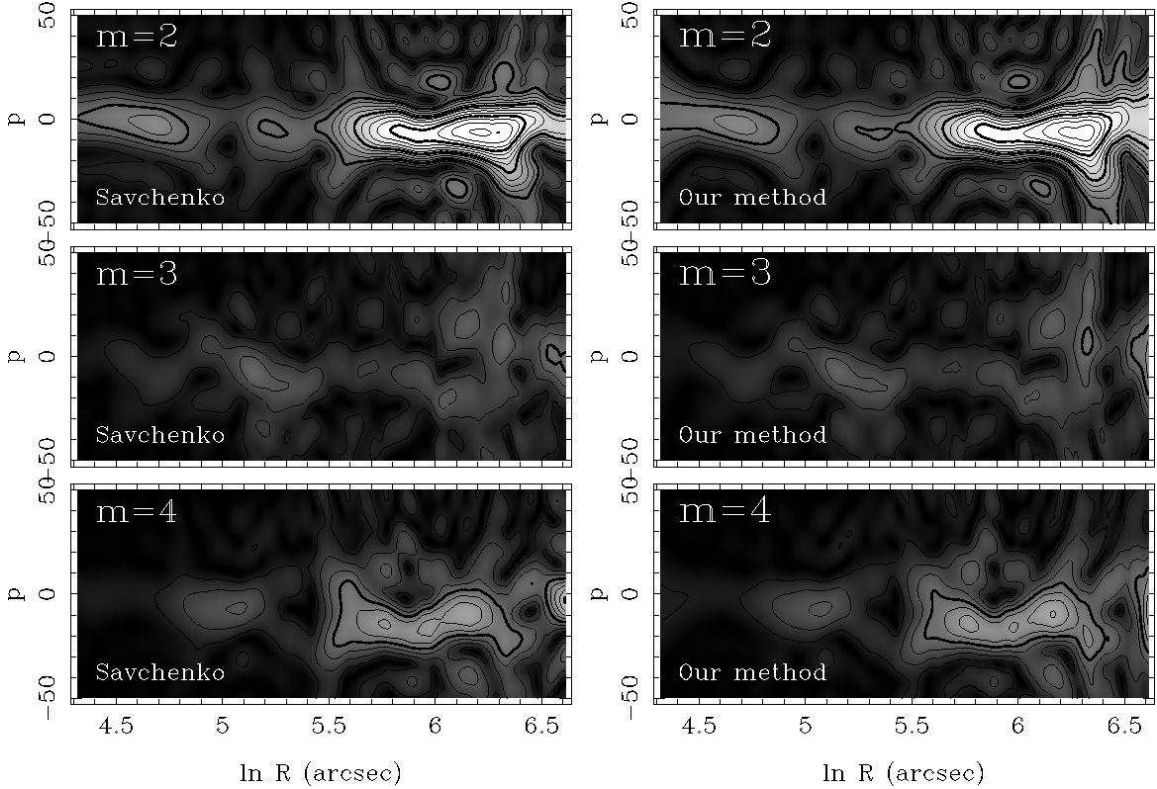}
\caption{Amplitudes of the 2D Fourier transform following the Savchenko method
(left) and correlation values for our method using a $2\pi$ window (right)
for $m=2, 3$, and 4. The $\Delta ln R$ for the calculations was
$u_{max}-u_{min}=0.35$ (see equation \ref{equation_2dft}). The $x$ axis is
the same as in Figure \ref{1dft}. The gray scale and the contour values are
the same for all of the plots. For $m=2$, note the low amplitudes for the 2D
Fourier transforms and for correlation values at $\ln R\sim$ 5.0, 5.5, and 6.4. For the
Savchenko method, we plot the amplitude of the $A(m,p)$ Fourier coefficient
as a function of $\ln R$ and frequency $p$, which is related to the
pitch angle $P$ as $\tan P=-m/p$. For our correlation method, we plot the
value of the correlation for each $\ln R$ and frequency $p$, namely, $C_{\rm m}
(\ln R, \theta, p)$ (equation \ref{corr_method_1}), maximized over $\theta$.}
\label{savchenko_ours}
\end{figure}

\begin{figure}
\centering
\includegraphics[width=6.0in]{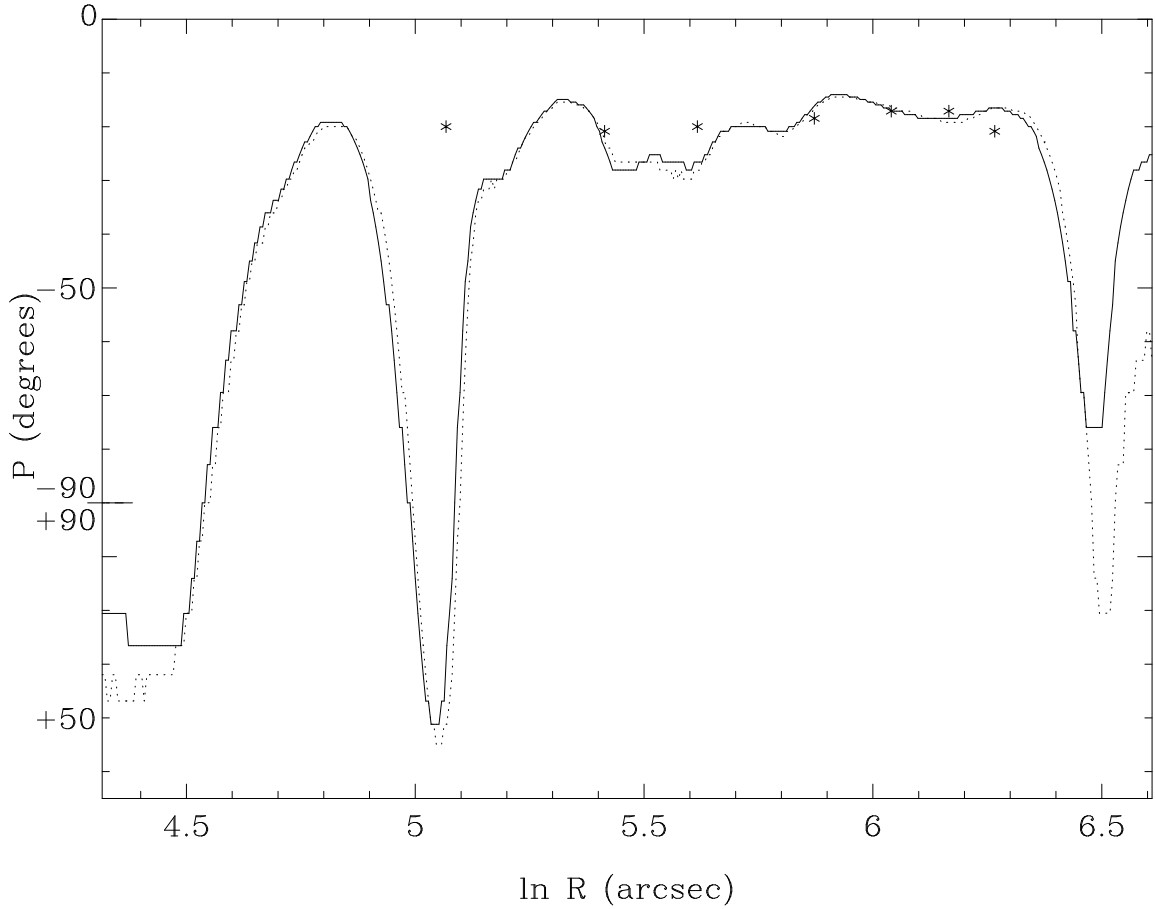}
\caption{Pitch angles for the $m=2$ component for the Savchenko method
(solid line), our correlation method (dotted line), and calculated as in
\cite{davisetal2012} (asterisks). Note the good agreement between the
three methods. The ``twist'' of the pitch angles around $\ln R=[5.0-5.1]$
occurs where $m=2$ has a minimum in amplitude (see Figure \ref{1dft}).
The results are not reliable for radii larger than $\ln R\sim6.4$.}
\label{pitch_angle_all_methods}
\end{figure}

The results of the application of the techniques discussed above are shown
for M81 in Figures \ref{galaxy_xy_lrt} to \ref{pitch_angle_all_methods}. In
Figure \ref{galaxy_xy_lrt}, we display the deprojected $8\mu$m image of M81
\citep[$PA=157^{\circ}$ and inclination $\omega=58^{\circ}$,][]{rc3} in 2
ways: in the plane of the galaxy, and in $(\ln R,\theta)$ space. Figure
\ref{1dft} displays the results of the application of equation
\ref{equation_1dft}. The amplitudes of the $m$ coefficients show that $m=2$
is the most prominent feature for this grand design galaxy. We can
distinguish two ``sets'' of $m=2$ structures: one from $R=150$ to $R=247$
arcsec and another from $R=247$ to $R=606$ arcsec. Figure
\ref{savchenko_ours} shows results for the 2D Fourier transform as a
function of radius \citep{savchenko2012} and for our correlation method
($2\pi$ window) for $\Delta\ln R=u_{max}-u_{min}=0.35$. Note that our
correlation method using $m$ synthetic spirals in a $2\pi$ window give
results very similar to those coming from a 2D Fourier transform. Finally,
in Figure \ref{pitch_angle_all_methods} we plot the estimation of pitch
angles for the $m=2$ component as a  function of radius for three different
techniques: \cite{savchenko2012}, \cite{davisetal2012} and our correlation
method ($2\pi$ window). Note that Savchenko's and our methods agree very
well for all radii. \cite{davisetal2012} only changed $\ln R_{min}$ in
their 2D Fourier transform. Definitely, the inner point calculated with the
Davis et al. method is affected by the outer structures.

\section{A new correlation method: circular windows in the $(\ln R, \theta)$ space}
\label{new_correlation}

\begin{figure}
\centering
\includegraphics[width=6.0in]{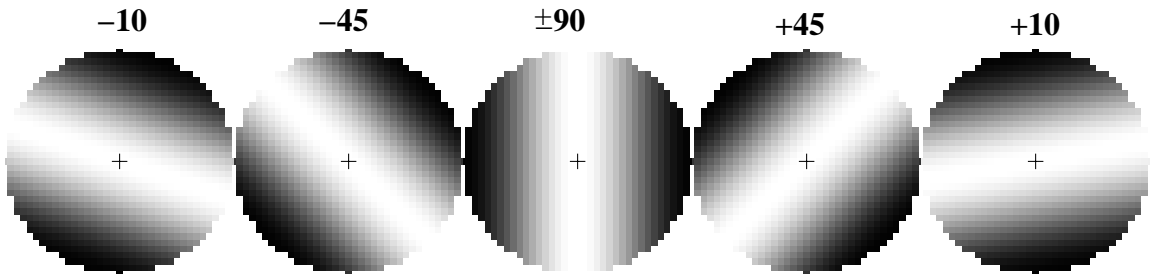}
\caption{Sine filters on circular windows. Horizontal direction: azimuth;
vertical direction: $\ln R$. White represents the maximum of the filter
values. Each filter is normalized, i.e., the sum of all values of a given
filter is always 1. Each filter is labeled with its pitch angle in degrees.
The plus sign indicates the central position $(\ln R_c, \theta_c)$. Filters
like these, with different diameters $D_w$ and with pitch angles $P_w$
varying from $-1^\circ$ to $-90^\circ$ and $+90^\circ$ to $+1^\circ$, are
used to calculate the cross correlation between the galaxy image at a given
$(\ln R, \theta)$ position and the filter. Higher correlations will be
achieved when the structure has the same pitch angle as the filter at that
$(\ln R, \theta)$ position. These correlations give a local estimate for
the pitch angle. Different windows diameters $D_w$ are used to stress the
difference between large and small scale structures.}
\label{circular_windows}
\end{figure}

Having checked our correlation method in a $2\pi$ window and getting
results in full agreement to those coming from the 2D Fourier transform, we
introduce a new method, now using circular windows in $(\ln R, \theta)$
space. By using these new windows, we no longer have a restriction on $m$
symmetry. Furthermore, by changing the diameter of the window, we can study
the distribution of the pitch angles at various scales. Inside each window,
we define a filter which is a sine function from 0 to $\pi$, i.e., we have
the maximum of the filter at its center, and the values falling to zero at
the borders of the window (Figure \ref{circular_windows}). So, for each
galaxy image, and for each position in $(\ln R, \theta)$ space, we
correlate the galaxy image with the filter function for different pitch
angles and window diameters. Let's describe the filter as $F=F(\ln R_c,
\theta_c, P_w, D_w)$, that is the circular filter centered at $(\ln R_c,
\theta_c)$, with pitch angle $P_w$, in a window of diameter $D_w$. We have
analysed our images for pitch angles $P_w$ from $-1^\circ$ to $-90^\circ$
and from $+90^\circ$ to $+1^\circ$, and for four different $D_w$ with
values 0.1, 0.2, 0.35 and 0.51 in units of $\ln R$. The algorithm first
transforms the deprojected $(x,y)$ image of a galaxy to a $(\ln R,
\theta)$ matrix. Our correlation method can be described as
\begin{equation}
	C(\ln R_c, \theta_c, P_w, D_w)={{\sum_{i=1,N} F(\ln R_c, \theta_c, P_w, D_w)
I(\ln R,\theta)}\over{\sum_{i=1,N} I(\ln R,\theta)}}
\end{equation}
where $\sum_{i=1,N}$ represents the summation over the $N=\pi(D_w/2)^2$
pixels on the circular window. $\sum_{i=1,N} I(\ln R,\theta)$ is a
normalization factor. The correlation will be higher when the pitch angle
of some structure centered at $(\ln R_c, \theta_c)$ coincides with the
pitch angle of the filter $F$. By changing the diameter of the window,
different spatial scales can be analyzed.

\section{Results for the locally determined pitch angles at different scales}
\label{results}

We deprojected the Spitzer $8\mu$m image of M81 using the values of
$PA=157^{\circ}$ and inclination $\omega=58^{\circ}$ \citep{rc3}. For M51,
$PA=170^{\circ}$ and inclination $\omega=20^{\circ}$ were taken, following
\cite{shettyetal2007}. The deprojected images were subjected to our cross
correlation method, using sine filters on circular windows in the $(\ln R,
\theta)$ space as described above.

\begin{figure}
\centering
\includegraphics[width=6.0in]{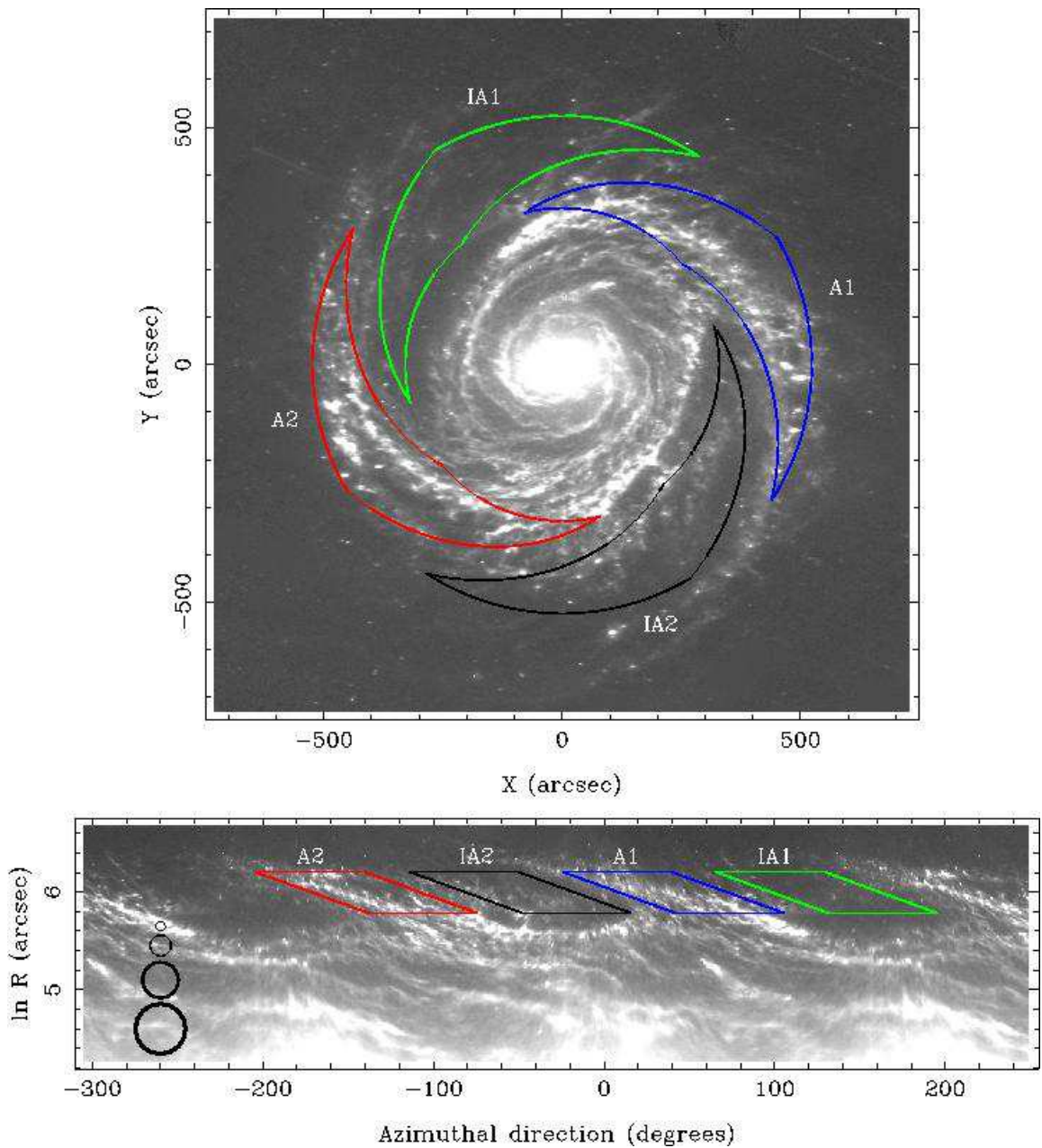}
\caption{As in Figure \ref{galaxy_xy_lrt}, M81 in $(x,y)$ and $(\ln R, \theta)$
representations. Here, we delineate the regions in which we add all of the
correlations for a given window diameter $D_w$ and for each pitch angle
$P_w$. The minimum and maximum radii are 330 and 525 arcsec, corresponding
to $\ln R=5.8$ and 6.26. The arm regions (blue and red) and interarms (green and
black) are marked. The circles drawn in the bottom panel represent
the four different window diameters $D_w$ we have used in this study; from
top to bottom, $D_w$= 0.1, 0.2, 0.35 and 0.51 in units of $\ln R$.} \label{regionsdelineated}
\end{figure}

In Figure \ref{regionsdelineated} we present once again M81 in $(x,y)$ and
$(\ln R, \theta)$, but now we delineate the areas in which we add all the
correlations calculated for a given window diameter for each pitch angle.
Four areas were chosen, two over the arms (blue and red), and two placed in
the interarm regions (green and black). The minimum and maximum radii for
the regions are 330 and 525 arcsec, or $\ln R=$5.8 and 6.26. As already
noted, we use four different window diameters $D_w$= 0.1, 0.2, 0.35, and
0.51 in units of $\ln R$. These circular windows are drawn in the bottom
panel of the figure, and represent the sizes of the features we are
measuring.

\begin{figure}
\centering
\includegraphics[width=6.0in]{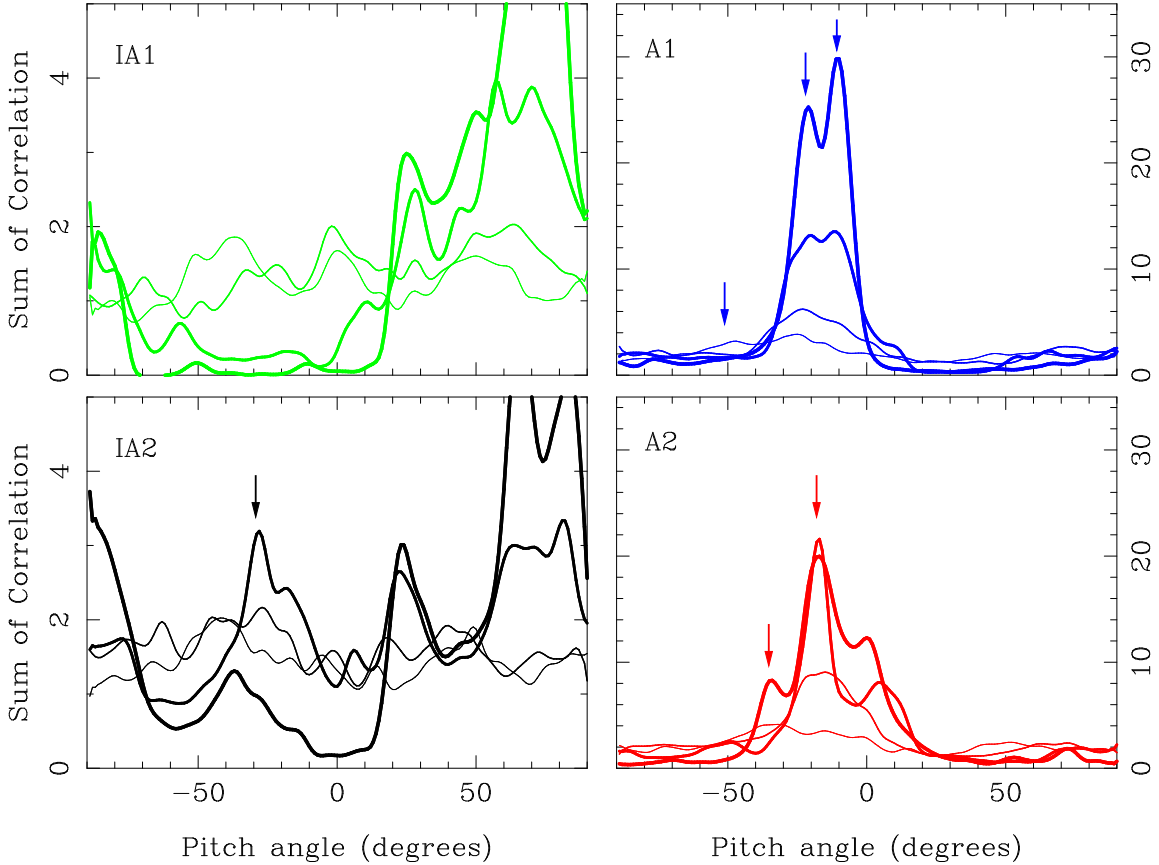}
\caption{Results for the sum of the correlations in the analysed areas for
M81. Green and black, interarm regions (left). Blue and red, arm regions
(right). Note the different scales in the left and right panels. The
thinnest line is for $D_w=0.1$, and the thickest one is for $D_w=0.51$.
Some peaks are marked with arrows in the panels. For the interarm regions
at small scales, the correlations are about the same for all pitch angles;
in the black region, one intermediate scale shows a pronounced peak at
$P\sim -30^\circ$. For larger scales, the correlations show contamination
from the arms (notice the high correlation for pitch angles $P\ge
+60^\circ$). For the arm regions at large scales, the blue area shows two
peaks of pitch angles around $P=-10^\circ$ and $P=-22^\circ$; the red area
has a strong peak at $P=-17^\circ$ and a secondary peak at $P\sim
-38^\circ$. For smaller scales, both arm regions (blue and red) show broad
peaks for $P\le -20^\circ, -25^\circ$.} \label{main_results}
\end{figure}

Figure \ref{main_results} shows the sum of correlations in each area for
the different window diameters, as a function of pitch angle (left panels,
interarm regions; right panels, arm regions). Thicker lines represent
larger window diameters and therefore represent longer spiral structures.
Thinner lines represent the smaller scale structures related to feathers.
For the interarm regions (green and black) and at smaller scales, we have a
broad range of pitch angles. For the black area, the distribution of pitch
angles has a peak at $P=-28^\circ$. For larger scales, the distribution is
affected by the main spiral arms at the azimuthal edge of the analysed
areas; pitch angles $P=90^\circ-P_{spiral}$ therefore get larger
correlations ($P_{spiral}$ is the pitch angle of the main spiral
structure).

For the arm regions in Figure \ref{main_results} (blue and red), the small
scales present a broad distribution of pitch angles and the large scales
show the main $m=2$ structure of this galaxy, which has a pitch angle of
$P\sim-17^\circ$ (see the marked peak for the red region in Figure
\ref{main_results}; see also Section \ref{1D2DFourier_and_correlation}).
The arm inside the blue area has two main components with $P\sim-22^\circ$
and $P\sim-10^\circ$. These two peaks can be recognized in the blue area in
Figure \ref{regionsdelineated}, bottom panel. At smaller radii, there is a
bright structure presenting a more open pitch angle. The structure at large
radius is similar to that in the red area.

\begin{figure}
\centering
\includegraphics[width=6.0in]{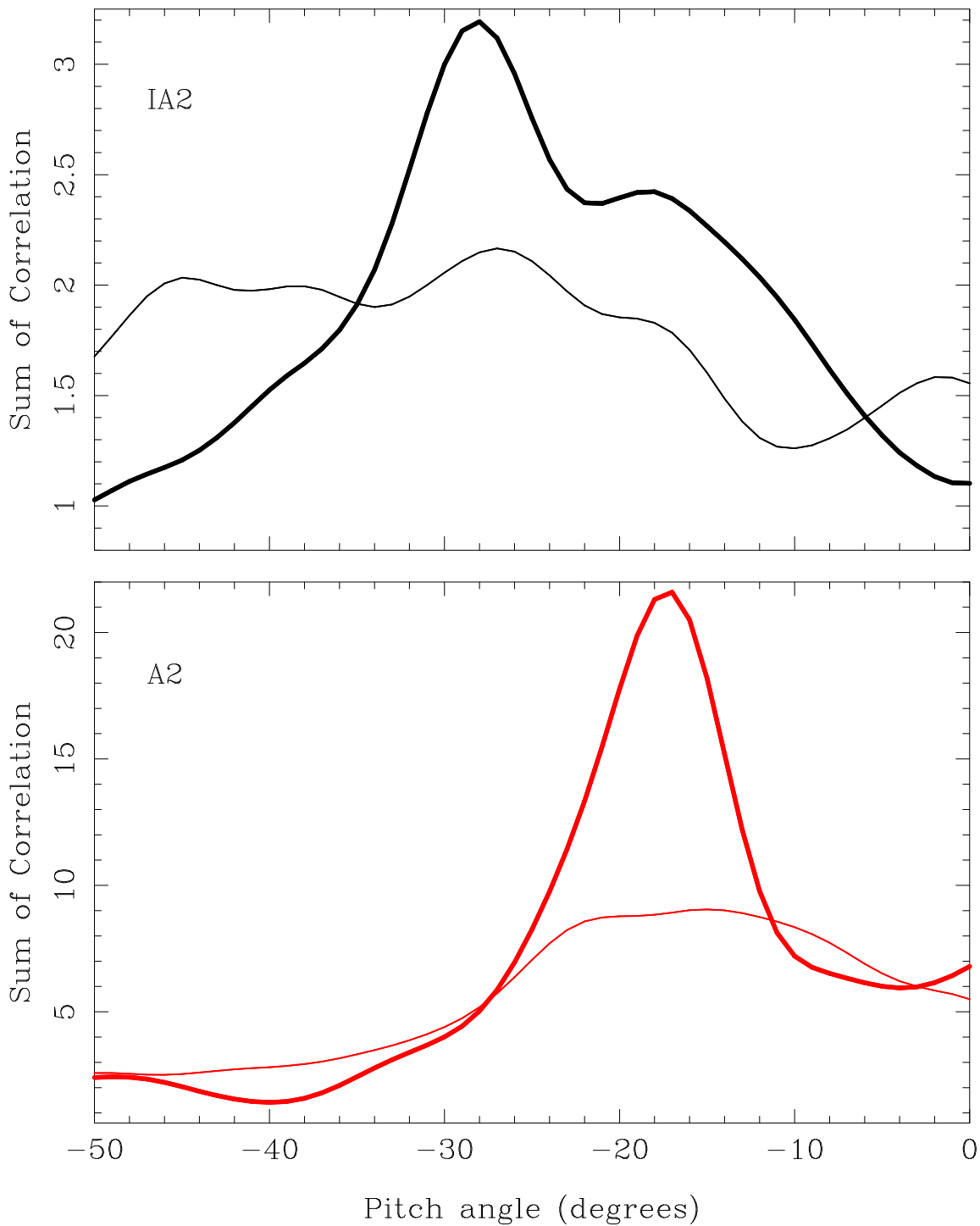}
\caption{Zoom in on the sum of correlations for the black (interarm) and red
(arm) regions in M81. Here we plot only the intermediary scales (thin line,
$D_w=0.2$, thick one, $D_w=0.35$). The arm region distribution of pitch
angles peaks at $P=-17^\circ$. For the interarm region, the sum of the
correlations points to more open structures, with a broader distribution in
pitch angles and a peak around $P=-28^\circ$.} \label{main_extra}
\end{figure}

In Figure \ref{main_extra}, we present a zoom of the sum of correlations
for one interarm (black) and one arm (red) region, for the two intermediate
scales $D_w$= 0.2 and 0.35. The arm peaks at a pitch angle around
$P=-17^\circ$, but for the interarm region, the sum of the correlations
points to a more open structure, with a broader distribution in pitch
angles and a peak around $P=-28^\circ$.

As discussed in the introduction, observations by \cite{lavigneetal2006}
and others show that feathers are more open compared to main spiral arms
(see their Figures 1 and 2 where they outline the feathers they detect by
eye in M51 and NGC 628). Numerical simulations \citep[e.g.,][]{kimkim14}
also show that sub-structures evolving from the main arms into the interarm
regions have larger pitch angles compared to the main spiral.

\begin{figure}
\centering
\includegraphics[width=6.0in]{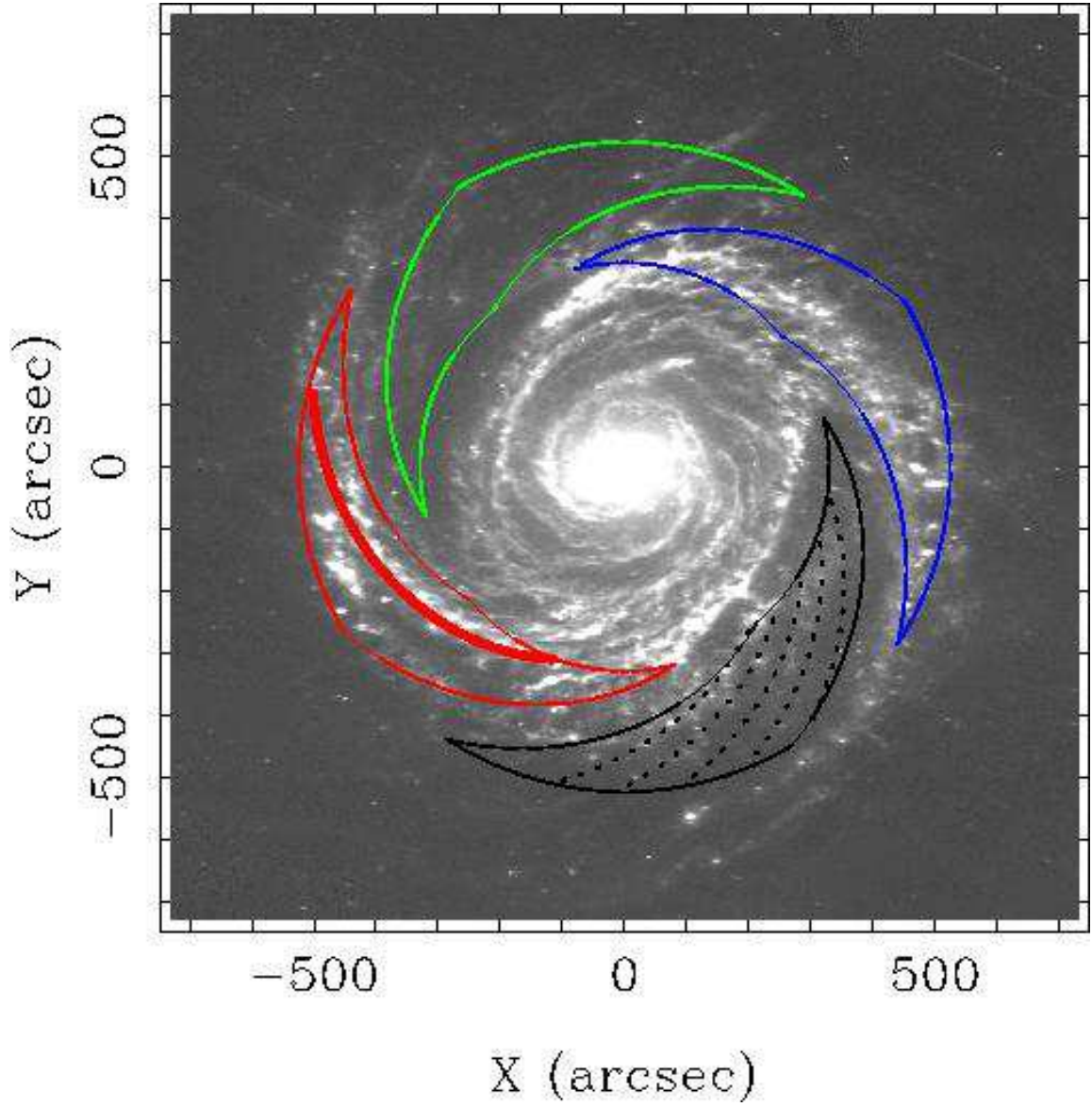}
\caption{Spiral arm tracings made from the results of Figure \ref{main_extra}
for M81. The thick solid red line represents the main spiral arm with its
measured pitch angle of $P=-17^\circ$. The dotted black lines represent
a pitch angle of $P=-28^\circ$. Note that these open interarm structures measured
here by the correlation technique match well to
the distribution of light.} \label{cartoon}
\end{figure}

In Figure \ref{cartoon}, we show a sketch of the results of Figure
\ref{main_extra}. The main spiral arm in the red region has $P=-17^\circ$
and is designated by a thick red curve. The dashed lines in the black
region have $P=-28^\circ$. These calculated values agree well with the
structures in the figure.

\begin{figure}
\centering
\includegraphics[width=6.0in]{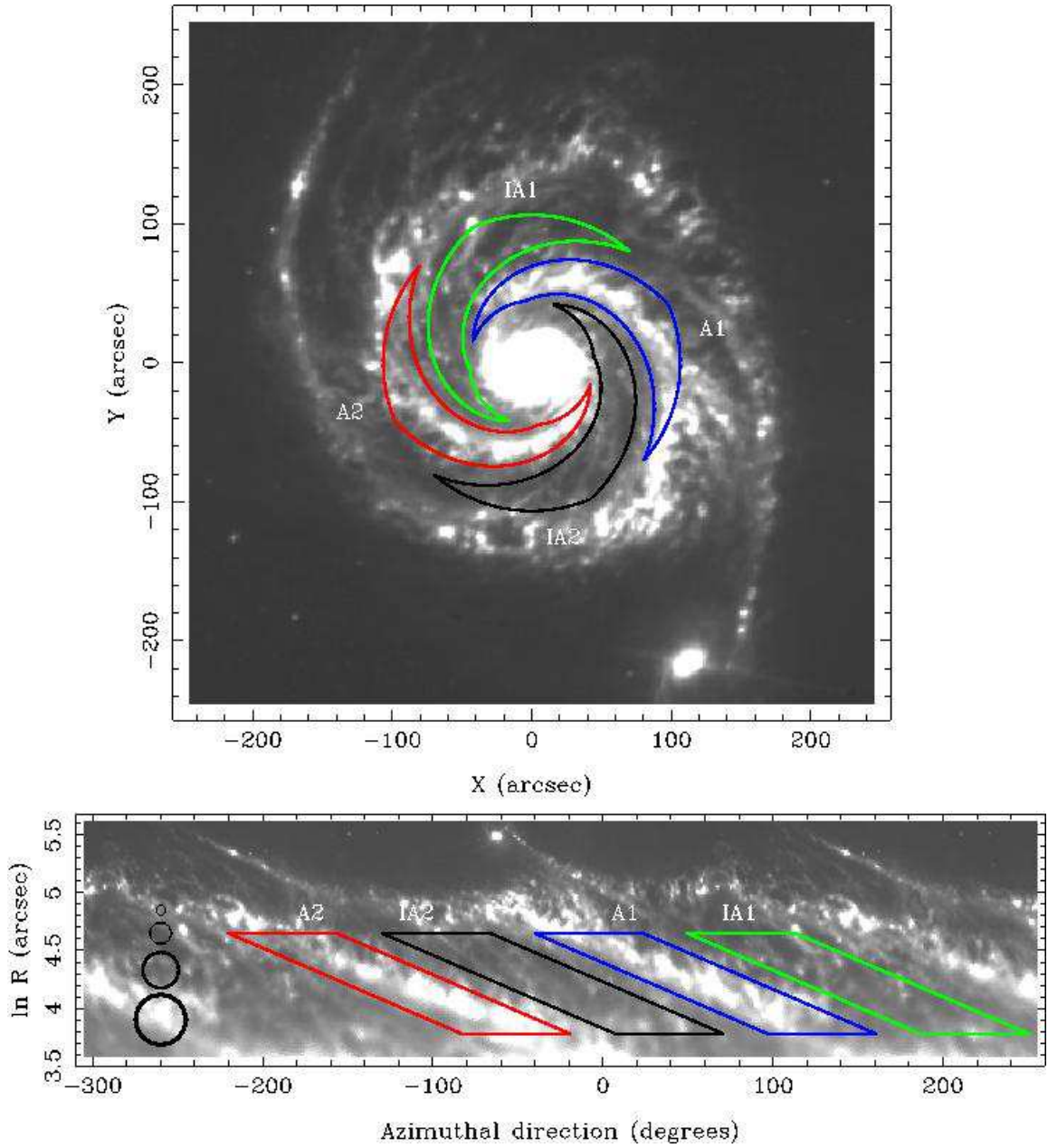}
\caption{As in Figure \ref{regionsdelineated}, but now for M51. Here, the
minimum and maximum radii are 44 and 106 arcsec, corresponding to $\ln
R=3.78$ and 4.66.} \label{regionsdelineatedm51}
\end{figure}

\begin{figure}
\centering
\includegraphics[width=6.0in]{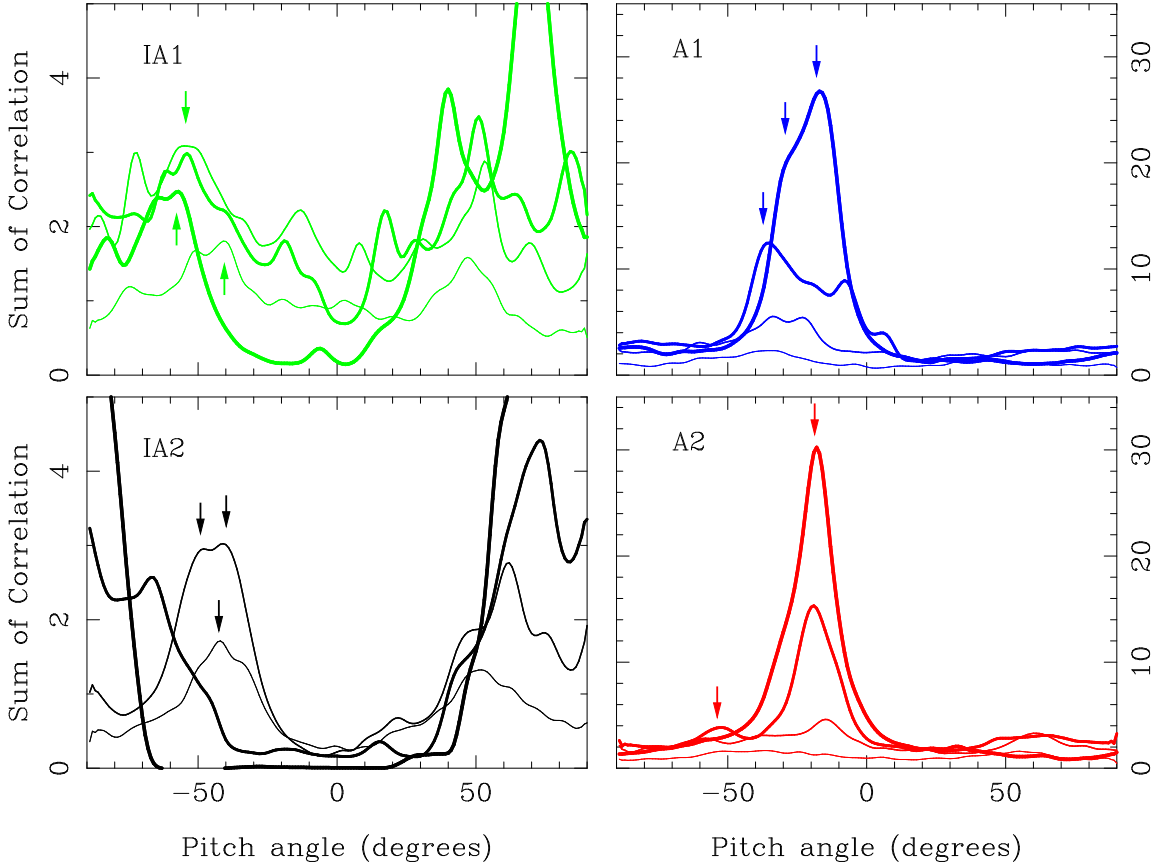}
\caption{As in Figure \ref{main_results}, but now for M51. Green and black are
interarm regions (left); blue and red are arm regions (right). Note the
different scales between left and right panels. The thinnest line is for
$D_w=0.1$, while the thickest one is for $D_w=0.51$. Some peaks are marked
with arrows in the panels. The distributions of pitch angles are
similar to those in M81. Here, both interarm regions (green and black)
have peaks for the pitch angles at $P\le -40^\circ$ covering a wide range
of scales. The
results for the arm regions (blue and red) at large scales represent the
main spiral arms, with $P\sim -19^\circ$. The blue region presents an
asymmetric main peak in the pitch angle distribution.
For smaller scales, both arm regions have broad
distributions of pitch angles, with larger values compared to that
of the main spiral arms.} \label{main_results_m51}
\end{figure}

\begin{figure}
\centering
\includegraphics[width=6.0in]{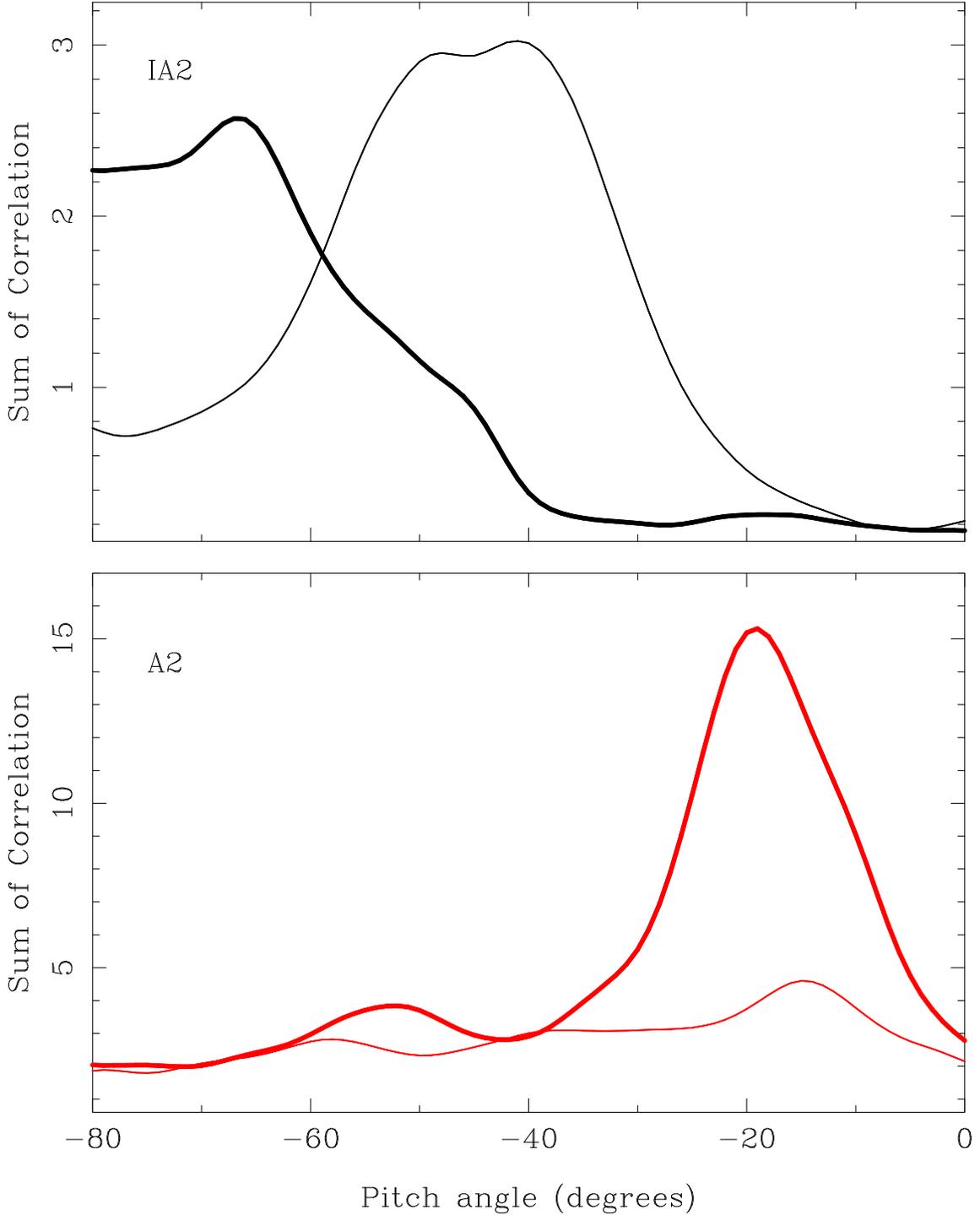}
\caption{Zoom in on the sum of correlations for the black (interarm) and red
(arm) regions in M51. Here we plot only the intermediary scales (thin line,
$D_w=0.2$; thick line, $D_w=0.35$). The arm peaks at a pitch angle around
$P=-19^\circ$, and presents a bump at $P\sim 52^\circ$. For the interarm
region, the sum of the correlations point to more open structures,
with a broader distribution in pitch angles and a peak around
$P=-65^\circ$. At the small scales plotted here, the interarm region shows a
broad distribution with a plateau from $P=-50^\circ$ to $P=-40^\circ$.}
\label{main_extra_m51}
\end{figure}

Figures \ref{regionsdelineatedm51} to \ref{main_extra_m51} are similar to
Figures \ref{regionsdelineated} to \ref{main_extra}, but now for M51. The
observed $(\ln R,\theta)$ map resembles the model at slow pattern speed in
\cite{kimkim14}. Figure \ref{regionsdelineatedm51} delineates the areas we
analysed for M51. We use minimum and maximum radii of 44 and 106 arcsec, or
$\ln R=$3.78 and 4.66. Figure \ref{main_results_m51} shows the sum of the
correlations for all regions. As for M81, the interarm regions on small
scales present a broad distribution of pitch angles up to a large value
(see the arrows in the left panels). The interarms on large scales have a
pitch angle distribution that is dominated by the main spiral arms, giving
large correlations for $P=90^\circ-P_{spiral}$. For the arm regions (right
panels), the spiral arm in the red region is more symmetric than the arm in
the blue region. The main peak for the spiral arm in the blue region is
quite asymmetric. Here again  we find secondary peaks for pitch angles
larger than the pitch angle of the main spiral arms.  In Figure
\ref{main_extra_m51} the pitch angle for the main spiral arm in the red
region is $P=-19^\circ$, with a secondary bump for $P\sim-52^\circ$ (bottom
panel). The structures in the interarm region have larger pitch angles,
$P\le -40^\circ$.

The pitch angles determined for the $m=2$ spiral arms are $P\sim -18^\circ$
for M81 and $P\sim -19^\circ$ for M51. For M81, \cite{kendalletal2008} used
Spitzer IRAC $3.6$ and $4.5\mu$m images to make an ``eye-ball'' fit giving
$P\sim -23^\circ$. Bash \& Kaufman (1986) estimated the pitch angles in
radio continuum maps. They noticed that the two main arms differ in pitch
angles, deriving $P\sim -17^\circ$ for the eastern arm and $P\sim
-23^\circ$ for the western arm. Similar numbers were determined with
bidimensional Fourier transforms by \cite{puerarietal2009} in a
multiwavelength study using GALEX and Spitzer IRAC images. For M51,
\cite{shettyetal2007} used the CO distribution to determine a pitch angle
of $P=-21.1^\circ$. More recently, \cite{huetal2013} used $P=-17.5^\circ$
from their models of M51. \cite{fletcheretal2011} report an average pitch
angle of $P=-20^\circ$. Evidently, many methods give similar results for
the pitch angles of the main spiral arms in these two galaxies. Here we
reproduce those values on large scales, but also find a broad range of
values up to larger pitch angles on small scales. There is a multiplicity
of spiral structures on different scales, as expected from gas flows in
gravitating, turbulent and shearing interstellar media.

\begin{acknowledgements}

I.P. acknowledges support from the Mexican foundation CONACyT and from the
University of the Witwatersrand.
\end{acknowledgements}

\clearpage

\clearpage

\clearpage

\clearpage

\clearpage

\clearpage

\clearpage

\clearpage

\clearpage

\clearpage

\clearpage

\clearpage

\end{document}